\documentclass[prl,twocolumn,amsmath,amssymb,superscriptaddress,unsortedaddress]{revtex4}

\usepackage{epsf}
\usepackage{graphicx}
\usepackage[usenames]{color}

\begin{document}

\title{Slater to Mott crossover in the metal to insulator transition of Nd$_2$Ir$_2$O$_7$}

\author{M.~Nakayama}
\affiliation{ISSP, University of Tokyo, Kashiwa, Chiba 277-8581, Japan}

\author{Takeshi Kondo}
\email{kondo1215@issp.u-tokyo.ac.jp}
\affiliation{ISSP, University of Tokyo, Kashiwa, Chiba 277-8581, Japan}

\author{Z.~Tian}
\affiliation{ISSP, University of Tokyo, Kashiwa, Chiba 277-8581, Japan}

\author{J.J.~Ishikawa}
\affiliation{ISSP, University of Tokyo, Kashiwa, Chiba 277-8581, Japan}

\author{M.~Halim}
\affiliation{ISSP, University of Tokyo, Kashiwa, Chiba 277-8581, Japan}

\author{C.~Bareille}
\affiliation{ISSP, University of Tokyo, Kashiwa, Chiba 277-8581, Japan}

\author{W.~Malaeb}
\affiliation{ISSP, University of Tokyo, Kashiwa, Chiba 277-8581, Japan}
\affiliation{Physics Department, Faculty of Science, Beirut Arab University, Beirut, Lebanon}

\author{K.~Kuroda}
\affiliation{ISSP, University of Tokyo, Kashiwa, Chiba 277-8581, Japan}

\author{T.~Tomita}
\affiliation{ISSP, University of Tokyo, Kashiwa, Chiba 277-8581, Japan}

\author{S.~Ideta}
\affiliation{UVSOR Facility, Institute for Molecular Science, Okazaki 444-8585, Japan}

\author{K.~Tanaka}
\affiliation{UVSOR Facility, Institute for Molecular Science, Okazaki 444-8585, Japan}

\author{M.~Matsunami}
\affiliation{Toyota Technological Institute, Nagoya 468-8511, Japan}

\author{S.~Kimura}
\affiliation{Graduate School of Frontier Biosciences, Osaka University, Suita, Osaka, 565-0871, Japan }

\author{N.~Inami}
\affiliation{Institute of Materials Structure Science, High Energy Accelerator Research Organization (KEK),  Tsukuba, Ibaraki 305-0801, Japan}

\author{K.~Ono}
\affiliation{Institute of Materials Structure Science, High Energy Accelerator Research Organization (KEK),  Tsukuba, Ibaraki 305-0801, Japan}

\author{H.~Kumigashira}
\affiliation{Institute of Materials Structure Science, High Energy Accelerator Research Organization (KEK),  Tsukuba, Ibaraki 305-0801, Japan}

\author{L.~Balents}
\affiliation{Kavli Institute for Theoretical Physics, Santa Barbara, California 93106, USA}

\author{S.~Nakatsuji}
\affiliation{ISSP, University of Tokyo, Kashiwa, Chiba 277-8581, Japan}
\affiliation{CREST, Japan Science and Technology Agency (JST), 4-1-8 Honcho Kawaguchi, Saitama 332-0012, Japan}

\author{S.~Shin}
\affiliation{ISSP, University of Tokyo, Kashiwa, Chiba 277-8581, Japan}

\date{\today}

\begin{abstract} We present an angle-resolved photoemission study of the electronic structure of the three-dimensional pyrochlore iridate Nd$_2$Ir$_2$O$_7$ through its magnetic metal-insulator transition.  Our data reveal that metallic Nd$_2$Ir$_2$O$_7$ has a quadratic band, touching the Fermi level at the $\Gamma$ point, similarly to that of Pr$_2$Ir$_2$O$_7$.  The Fermi node state is, therefore, a common feature of the metallic phase of the pyrochlore iridates.  Upon cooling below the transition temperature, this compound exhibits a gap opening with an energy shift of quasiparticle peaks like a band gap insulator.  The quasiparticle peaks are strongly suppressed, however, with further decrease of temperature, and eventually vanish at the lowest temperature, leaving a non-dispersive flat band lacking long-lived electrons. We thereby identify a remarkable crossover from Slater to Mott insulators with decreasing temperature.  These observations explain the puzzling absence of Weyl points in this material, despite its proximity to the zero temperature metal-insulator transition.  \end{abstract}


\maketitle

The $5d$ iridium oxides (iridates), having comparable scales for their kinetic energy, coulomb interaction and spin-orbit coupling, provide an excellent platform to study new types of strongly correlated phenomena \cite{Deller:2009dg,Kim:2008gi,Kim:2014ku,Pesin:2010dg,Yang:2010dv,WitczakKrempa:2012cd,Wan:2011hia,Nakatsuji:2011gg,Jackeli:2009hz,Chaloupka:2010gi,You:2012cv}.  Amongst them, the pyrochlore iridates ($Ln$$_2$Ir$_2$O$_7$, where $Ln$ is a lanthanide),  endowed with frustrated geometry and cubic symmetry, have a particularly fascinating phase diagram. Pr$_2$Ir$_2$O$_7$, with the smallest $Ln$-ion, is a metallic spin-liquid \cite{Nakatsuji:2006fc,Tokiwa:2014ik,MacLaughlin:2010kq} and exhibits an anomalous Hall effect \cite{Nakatsuji_Nature,Balicas:2011fw}.   For $Ln$-ions with larger ionic radius, an antiferromagnetically ordered insulating phase appears at low temperature.
 
Theoretically, topological band structures have been ascribed to the $Ln$$_2$Ir$_2$O$_7$ series \cite{Yang:2014cx,Pesin:2010dg,WitczakKrempa:2014hz,Wan:2011hia,Chen:2012ek}. 
The metallic phase is predicted to exhibit  quadratically dispersing conduction and valence bands touching at the $\Gamma$ point close to the Fermi level ($E_{\rm F}$) \cite{Savary:2014ct,Moon:2013ha}.   This structure has been recently identified by angle-resolved photoemission spectroscopy (ARPES) in Pr$_2$Ir$_2$O$_7$ \cite{kondo2015quadratic}. Theory predicts that such a quadratic Fermi node state may be converted into various topological states such as a topological insulator or a Weyl semimetal by appropriate symmetry breaking \cite{Yang:2014cx,Pesin:2010dg,WitczakKrempa:2014hz,Wan:2011hia,Chen:2012ek}. % Notably, HgTe, known as the first discovered topological insulator, also has a quadratic Fermi node \cite{Liu:2015ji,Brune:2011hi,Zaheer:2013hc}, so that metallic $Ln$$_2$Ir$_2$O$_7$ could be viewed as a strongly correlated analog of HgTe \cite{kondo2015quadratic}. 

Antiferromagnetism in these materials is of Ising type, consisting of an ``All-In-All-Out'' (AIAO) configuration of Ir moments on alternating tetrahedra \cite{muSR_Tomiyasu,muSR_Guo,Arima,DomainWall_Shen}.  This can be considered an ``octupolar'' spin order which breaks time-reversal but preserves cubic symmetry, and does not enlarge the unit cell \cite{Arima:2013ga}.  The Ising nature implies two types of domains, which have recently been shown, in agreement with theoretical predictions \cite{Yamaji:2014fa}, to be separated by metallic domain walls \cite{Tian_NP,Ueda:2015dt},  which have been imaged by microwave impedence microscopy in the magnetic state of Nd$_2$Ir$_2$O$_7$ \cite{DomainWall_Shen}.  Early density functional studies predicted the magnetic state to be a Weyl semimetal \cite{Wan:2011hia}, and general arguments imply that if a quasiparticle picture applies at low energy in the antiferromagnetic phase, and the magnetic ordering is weak, it {\em must} exhibit Weyl points and cannot have a true gap.  Nevertheless, optical \cite{Ueda:2012ee} and transport \cite{Tian_NP} measurements indicate a gapped insulating ground state for Nd$_2$Ir$_2$O$_7$, despite its low antiferromagnetic/Metal-Insulator (MI) transition temperature $T_{\rm MI}\approx 30$K and proximity to metallic Pr$_2$Ir$_2$O$_7$.   This begs the question of whether the weakness of the order, the quasiparticle assumption, or both, break down in this system.  More generally, we seek to understand the influence of the MI transition on the conduction electrons.

In this letter, we use ARPES to investigate the evolution of the electronic structure through the MI transition in Nd$_2$Ir$_2$O$_7$, which is the most suitable member of the series for such study because its low $T_c$ minimizes thermal broadening.  Although the layered iridates have been extensively studied by photoemission \cite{Na2IrO3_Damascelli,Kim:2008gi,King:2013ea,Nie:2015jt,Kim:2014hx,kim_NP,Tamai_Sr2IrO4,Tamai_Sr3Ir2O7}, ours is the first study across a MI transition in any iridate, since the latter occurs only in the pyrochlores, for which preparation of a proper crystal surface is difficult.  Having overcome this challenge, we are able to directly measure both the single particle excitations of the metallic and insulating phases.  In the former, we observe features very similar to those reported for Pr$_2$Ir$_2$O$_7$ \cite{kondo2015quadratic}, indicating a quadratic band touching at the $\Gamma$ point at the Fermi energy, with a valence band that is greatly narrowed (width $\sim$40 meV) in comparison to density functional calculations.  Below $T_{\rm MI}$, the band touching is removed and a gap develops.  The magnitude of the gap, $\sim 40$ meV, is similar to that observed in optics, and comparable to the observed bandwidth, consistent with {\em strong} magnetic order.  These features might be understood from a Slater-type quasiparticle picture.  However, we also observe that the quasiparticle peak is strongly suppressed on cooling below $T_{\rm MI}$, and the valence band becomes essentially completely flat at the lowest temperature, which indicate a Mott-type insulating state with correlation-derived localization.  We conclude that Nd$_2$Ir$_2$O$_7$ displays a dramatic Slater to Mott crossover with reducing temperature.  % The Mott-like low temperature state is probably incompatible with the formation of non-trivial band topology and Weyl fermions. 
This implies that Weyl fermions, if they exist, may do so only in a narrow region of temperature slightly below  $T_{\rm MI}$, in which the order is indeed weak and quasiparticles can survive. 

% Because of its low transition temperature, correlation-derived thermal broadening is minimized \cite{Tomiyasu:2012fx}, making it the most suitable member of the series to study the MI transition.   In contrast to the layered iridates, which have been intensively studied by ARPES\cite{Na2IrO3_Damascelli,Kim:2008gi,King:2013ea,Nie:2015jt,Kim:2014hx,kim_NP,Tamai_Sr2IrO4,Tamai_Sr3Ir2O7}, no three-dimensional MI transition has been probed  While the gap feature of insulating Nd$_2$Ir$_2$O$_7$ was previously discussed with the optical conductivity \cite{Ueda:2012ee}, a lack of momentum-resolved information prevents one from going beyond speculation about it.  Besides, none of ARPES results on the MI transition of three dimensional iridates has been reported so far owing to a difficulty in preparing a proper crystal surface. The direct observation of band dispersion are, therefore, strongly desired to clarify the electronic properties of Nd$_2$Ir$_2$O$_7$, and to explore the prospects for a Weyl semimetal state in pyrochlore iridates \cite{Pesin:2010dg,Wan:2011hia,WitczakKrempa:2012cd,Moon:2013ha}. 

Single crystals of Nd$_2$Ir$_2$O$_7$ with $\sim$1 mm$^3$ in size were grown with a flux method. The cleavage surface measured by ARPES is the (111) plane. The ARPES experiments were performed at BL7U of UVSOR facility with a MBS A-1 analyzer ($h\nu = 8\sim 18$ eV) \cite{Kimura:2010ef}, BL28A of Photon Factory in KEK with a Scienta SES2002 analyzer ($h\nu = 39\sim 60$ eV),  and 1$^3$ beamline in BESSY-II with a Scienta R4000 analyzer ($h\nu = 50\sim 60$ eV). 
The overall energy resolution in  ARPES was set to $\sim$15 meV, and the lowest achievable temperature was 1K.

As previously reported, the transition temperature $T_{\rm MI}$ in $Ln$$_2$Ir$_2$O$_7$ \cite{Matsuhira:2011ku} is controlled by the $Ln$-ion size \cite{Ueda:2015cb}, the pressure \cite{Ueda:2015cb,Sakata:2011ie}, and the off-stoichiometry \cite{Ishikawa:2012gt}. 
 We have selected three pieces of Nd$_2$Ir$_2$O$_7$ crystals with different transition temperatures to investigate the variation of the MI transition with small changes in stoichiometry \cite{Ishikawa:2012gt,Tian_NP,MagTrans_Ueda}. We confirmed, with an electron-probe microanalysis (EPMA), 
a slight deviation from stoichiometry 
in the Ir/Nd ratio of approximately 1$\%$ and 2$\%$ for the single crystals with zero-field $T_{\rm MI}$ of $\sim$25K and $\sim$20K, respectively.
 Figure 1 shows the resistivity, $\rho(T)$, of the crystals we used for ARPES; 
note that we retrieved the crystal-piece after the ARPES experiment, and measured the 
resistivity of exactly the same piece to properly compare the ARPES and resistivity results. The temperature derivative of  $\rho(T)$, ${d\rho (T)/d T}$,  (inset panel) enables us  to estimate the value of $T_{\rm MI}$ from the onset of its reduction. 
As marked by arrows, different transition temperatures $T_{\rm MI}$ of $\sim$19K, $\sim$25K, and $\sim$36K were estimated for the three samples, which are thus labeled as MI19K, MI25K, and MI36K for the rest of the paper.

% \\\\\\\\\\\\\\\\\\\\\\\\\\\\\\\\

%We performed ARPES measurement using same batch samples as shown in Fig.1. Arrows in its figure show the MI transition temperature determined by ARPES. We defined MI transition temperature as the time when the gap begins to change. To clear the change of resistivity, Inset figure shows the differentiation of resistivity. Below transition temperature, $\rho(T)$ shows insulating (negative $d\rho/dT$) behavior. From these figure, we can vaify the correspondence of transition temperature between ARPES and resistivity. These resistivity data are appeared to change their behavior around the lower temperature than their MI transition temperature. According to the neutron research\cite{Tomiyasu:2012fx}, MI transition of Nd$_2$Ir$_2$O$_7$ is considered that Ir atoms induce MI transition by forming all-in all-out magnetic structure. After that, Nd moment forms all-in all-out structure at the more lower temperature. This Nd moment cooperate with Ir magnetic structure and is speculated to accelerate the insulating behavior. 

 %%%%%%%%%%%%%%%%%%%%%%
\begin{figure} \includegraphics[width=2.3in]{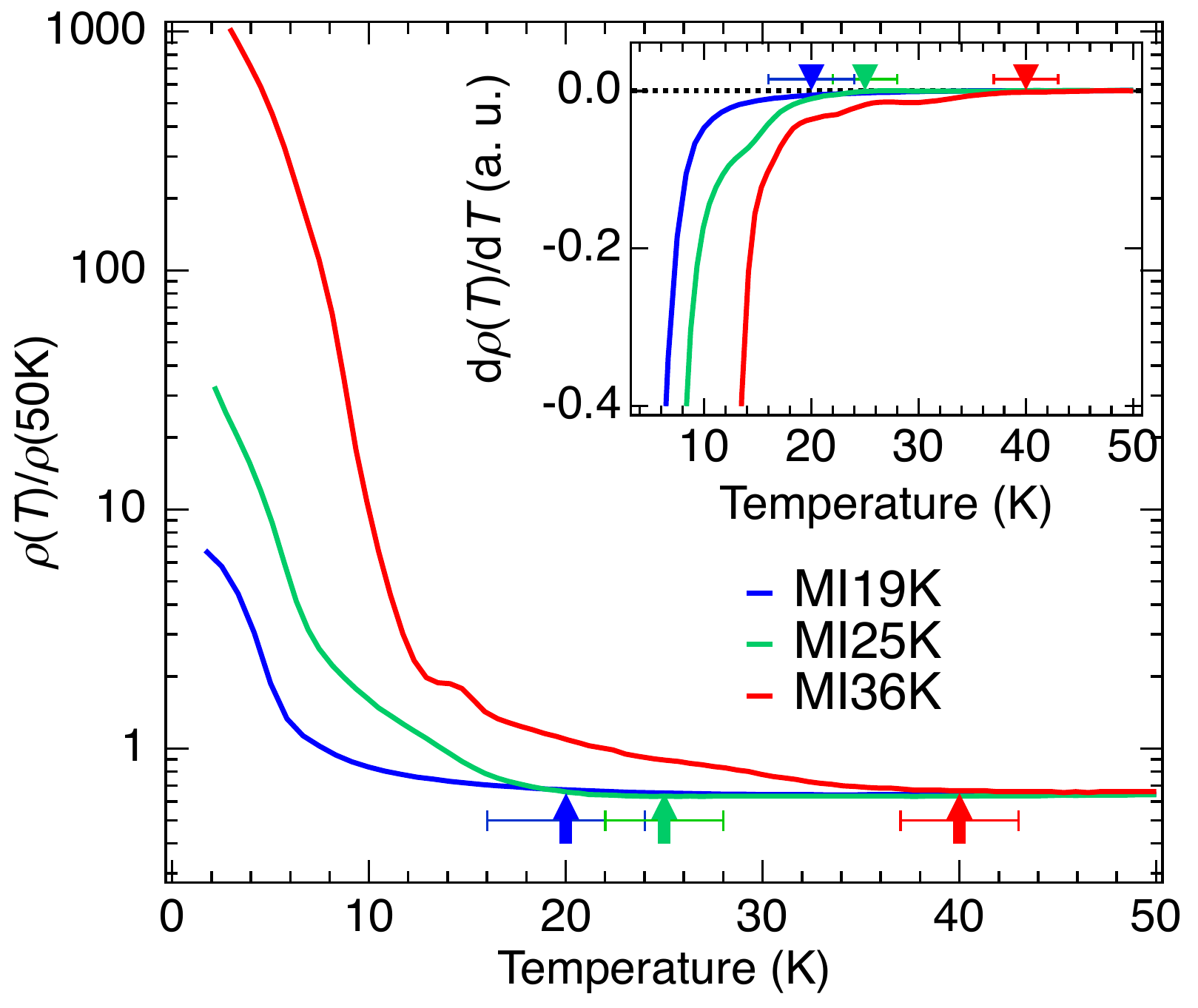} 
\caption{(color online) Temperature dependence of the resistivity, ${\rho (T)}$, for  Nd$_{2}$Ir$_{2}$O$_{7}$ crystals (MI19K, MI25K, and MI36K) we used for ARPES measurements. It is normalized to the intensity at $T$=50K. Inset panel plots the temperature derivative of the resistivity, $d\rho (T)/dT$. The transition temperature ($T_{\rm MI}$) estimated is marked by an arrow.}
\label{fig0_R}
\end{figure}
%%%%%%%%%%%%%%%%%%%%%%

In Fig. 2, we examine the band structure in the metallic phase.
Figure 2(c1) plots the typical ARPES spectra (energy distribution curves: EDCs) obtained at ($k_x, k_y$)=(0,0) with low-energy photons $(9.0{\rm{eV}} \le h\nu  \le 11.5{\rm{eV}})$ corresponding to  $k_z$ (or $k_{\rm (111)}$) values in the 1st Brillouin zone. Small but sharp quasiparticle peaks are observed for all of the photon energies as marked by arrows in Fig. 2(c1). We find that the quasiparticle peak approaches $E_{\rm F}$ with increasing photon energies and moves away again after getting closest to it at $h\nu $=10.5eV.  In Fig. 2(c2), the EDCs are symmetrized about $E_{\rm F}$ to remove the effect of Fermi cut-off \cite{kondo2015quadratic,SM}.
We found that the gapped spectra with two peaks merge to one peak at 10.5eV, thus the band touching occurs in Nd$_{2}$Ir$_{2}$O$_{7}$ at the same photon energy as in Pr$_{2}$Ir$_{2}$O$_{7}$ \cite{kondo2015quadratic}. 
To validate this further, we also used higher photon energies reaching the 3rd Brillouin zone (green circles in Fig. 2(a)), and reproduced the Fermi node again at  $\Gamma$ ($h\nu $=53eV) as shown in Fig. 2(d) \cite{SM}.

While ARPES is a technique to observe the occupied band structure, one can visualize the unoccupied states slightly above $E_{\rm F}$ by raising the sample temperature. Figure 2(b) demonstrate such an ARPES image along a $k_x$ cut across $\Gamma$.   Here the intensities are divided by the Fermi function at the measured temperature  ($T$=75K) to properly reveal the spectra above $E_{\rm F}$. The spectrum becomes broad due to the short lifetime characteristic of strongly correlated systems at high temperatures, so it is not possible to detect the quasiparticle peaks in the unoccupied side.  Nonetheless, significant intensities, indicative of the theoretically predicted conduction-band, are visible (a black arrow in Fig.2(b)). 

%%%%%%%%%%%%%%%%%%%%%%
\begin{figure} \includegraphics[width=2.8in]{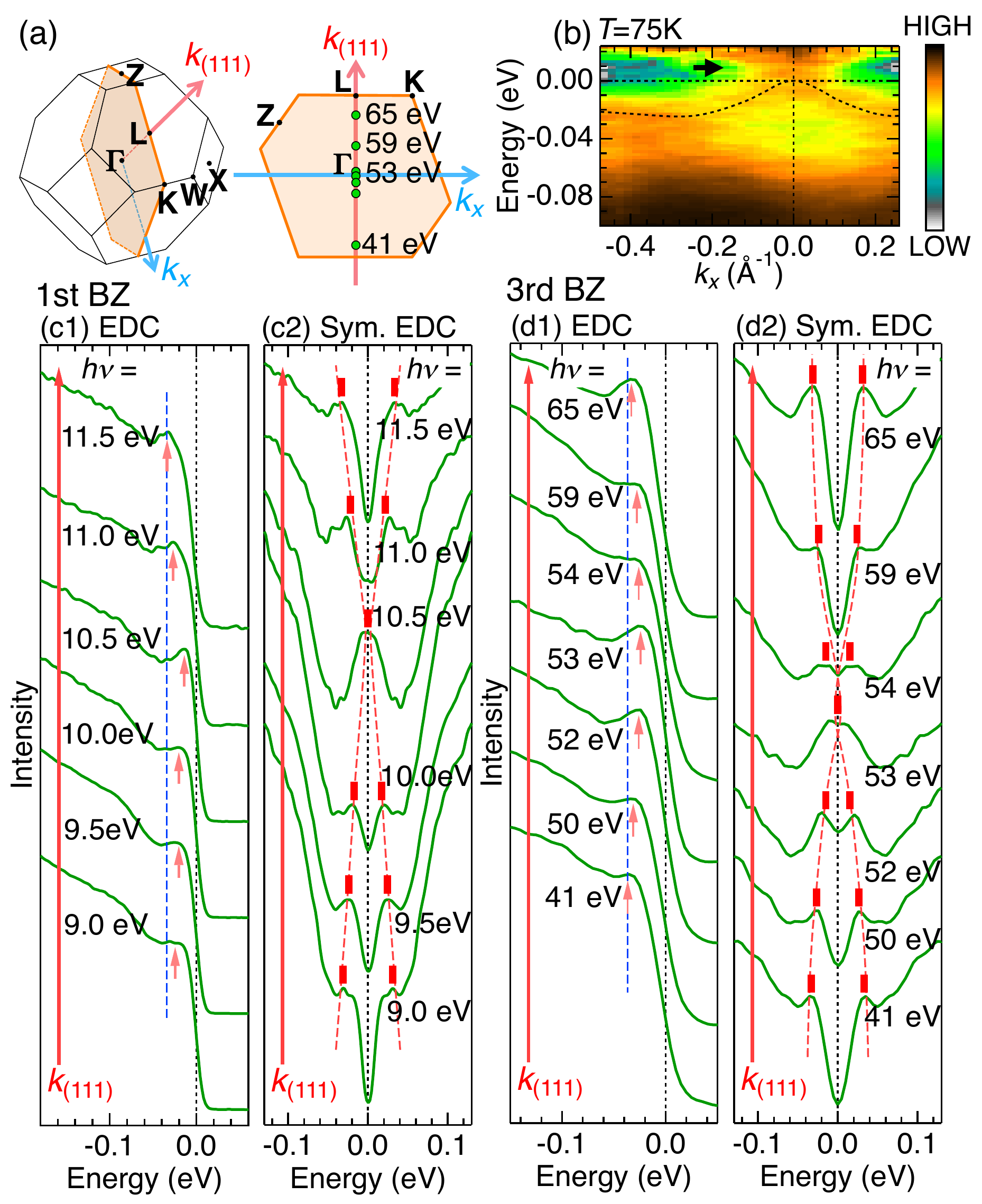} 
\caption{(color online)  (a)  Brillouin zone for Nd$_2$Ir$_2$O$_7$. (b) Band dispersion map crossing $\Gamma$, divided by the Fermi function at the measured temperature ($T$=75K). The arrow indicates the intensities implying an expected conduction-band. 
    EDCs ($T$=15K)  at ($k_x, k_y$)=(0,0) measured with low-energy photons (c1) and high-energy photons (d1), corresponding to $k_{(111)}$s in the 1st and 3rd Brillouin zone, respectively.
(c2,d2) The same data as in (c1) and (d2), respectively, but symmetrized about $E_{\rm F}$. 
Arrows and bars mark peaks in the  spectra. } 
\label{fig1}
\end{figure}
%%%%%%%%%%%%%%%%%%%%%%

Intriguingly the band width of Nd$_2$Ir$_2$O$_7$ is found to be extremely narrow, of order $\sim$40 meV on the occupied side, which is much less than expectated from DFT calculations.  While a band narrowing is also reported for the other iridates such as Na$_2$IrO$_3$ \cite{Na2IrO3_Damascelli}, Sr$_2$IrO$_4$ \cite{Kim:2008gi}, Sr$_3$Ir$_2$O$_7$ \cite{King:2013ea} and SrIrO$_3$ \cite{Nie:2015jt}, it seems to be comparable or even more significant in the pyrochlore iridates, consistent with   DMFT calculations \cite{Shinaoka:2015cg}. % which also reproduce the extreme band narrowing with a realistic value of on-site Coulomb repulsion U, as well as the smearing out of lower valence bands, compatible with out data\cite{Shinaoka:2015cg}. 
Furthermore, we detect a peak-dip-hump structure in the spectra, as is often observed in strongly correlated systems. These results are consistent with those of Pr$_2$Ir$_2$O$_7$ \cite{kondo2015quadratic}.  The observations in both materials are consistent with a picture of the metallic state as a highly renormalized Fermi liquid \footnote{A ``weak'' non-Fermi liquid which still possesses quasiparticle peaks would also be compatible \cite{Moon:2013ha}.}, with small quasiparticle weight $Z$ and large effective mass/bandwidth narrowing, for which a commensuraturely low coherence scale $\epsilon_{\rm coh}$ of tens of meV is expected.  The latter would naturally explain the strong broadening of bands at energies below $\sim -0.1$eV \cite{Shinaoka:2015cg}.

%%%%%%%%%%%%%%%%%%%%%%
\begin{figure}
\includegraphics[width=3.1in]{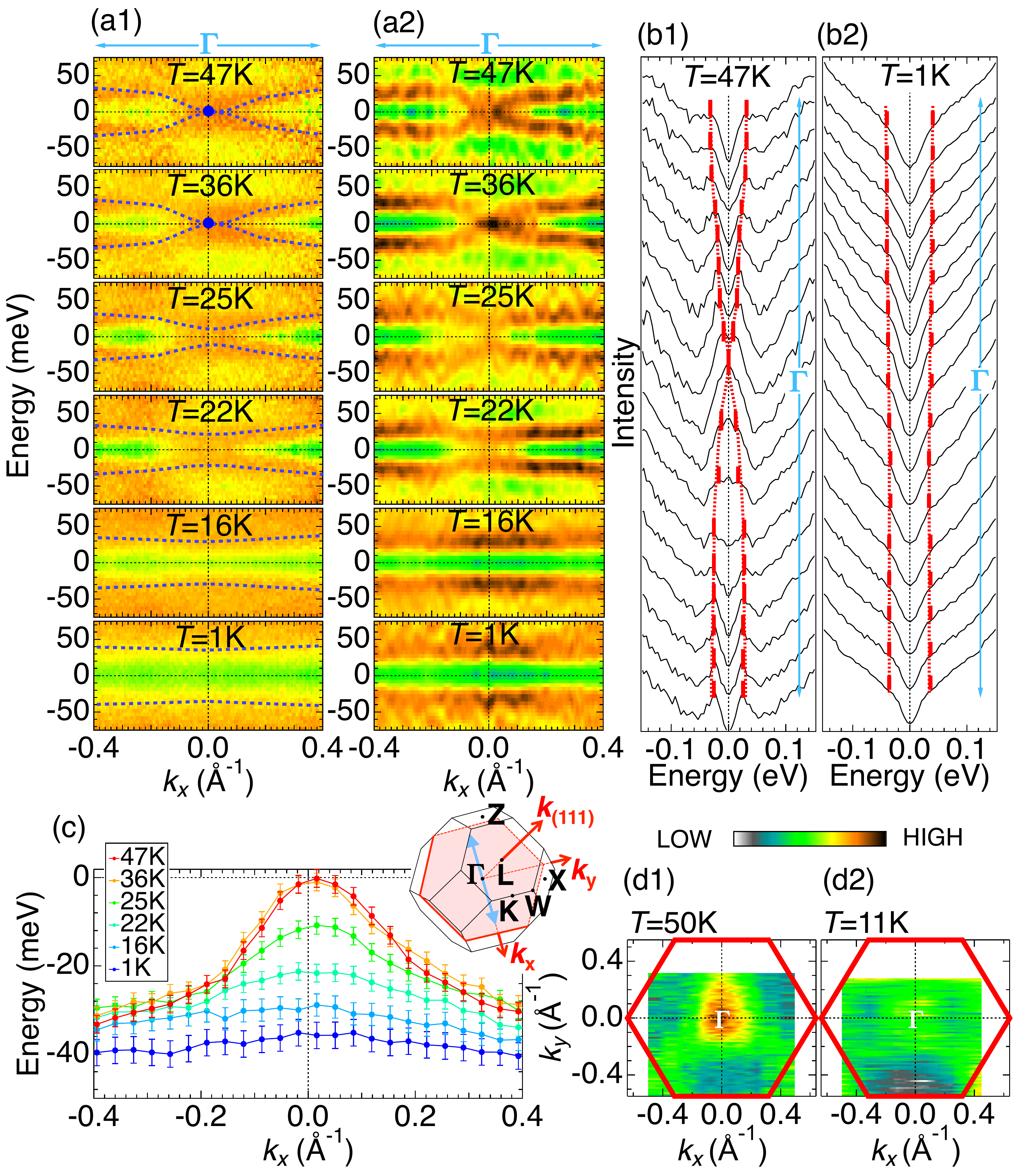}
\caption{(color online)  (a1) Band dispersion map across $\Gamma$ ($h\nu$=53eV; a light blue arrow in the inset of (c)) measured at various temperatures. The images are symmetrized about $E_{\rm F}$. Blue dashed curves indicate the obtained band dispersions. 
 (a2) Second derivative of the images in (a1).  
 (b1,b2) Spectra extracted from (a1) for the metallic phase ($T$=47K) and the insulating phase ($T$=1K), respectively.  
 (c) Temperature dependence of the band dispersion determined from the spectral peaks or shoulders (red bars in (b1) and (b2)). 
 (d1,d2) Spectral intensities at $E_{\rm F}$  along a momentum sheet crossing $\Gamma$ (red region  in the inset of (c)), 
 measured for the metallic phase ($T=50K$) and the insulating phase ($T=$11K), respectively. }
\label{fig2}
\end{figure}
%%%%%%%%%%%%%%%%%%%%%%

%%%%%%%%%%%%%%%%%%%%%%
\begin{figure}
\includegraphics[width=2.8in]{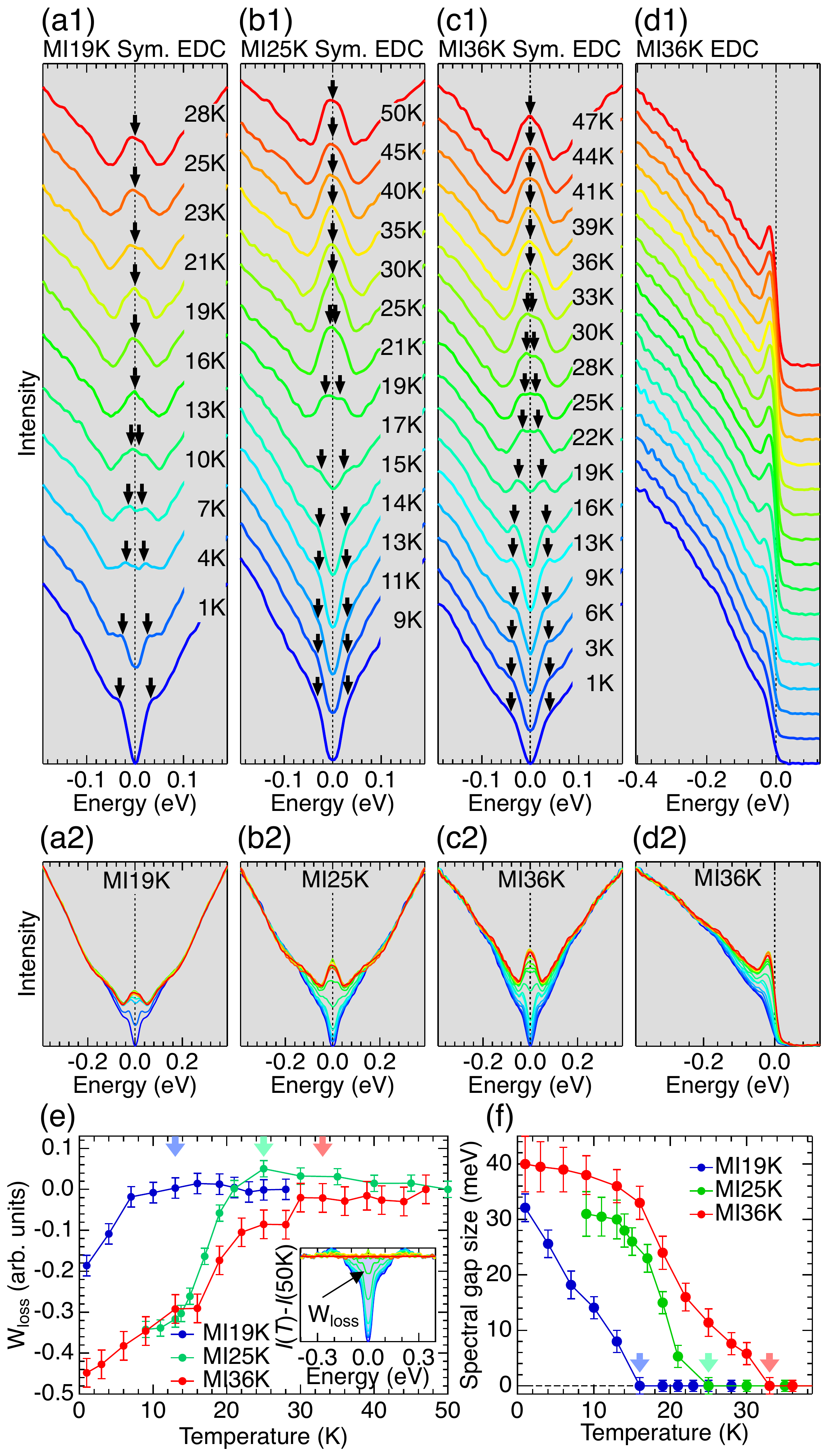}
\caption{(color online)  (a1-c1) Temperature evolution of symmetrized EDCs for three samples (MI19K, MI25K, and MI36K) measured at the $\Gamma$ point.  (d1) The same data as in (c1), plotted without symmetrization. (a2-d2) The same data as in (a1-c1), but without an offset. 
(e) Temperature dependence of spectral wight loss near $E_{\rm F}$ ($W_{\rm loss}$), which
is determined to be a negative area in the difference spectra as demonstrated in the inset.  
 (f) Temperature dependence of the magnitude of energy gap (arrows in (a1-c1)) estimated from the spectral peak positions in (a1-c1).}
\label{fig3}
\end{figure}
%%%%%%%%%%%%%%%%%%%%%%

We now turn to the MI transition.  In Fig. 3, we examine the temperature evolution of band dispersion through $T_{\rm MI}$, measured along a momentum cut across $\Gamma$ (a light blue arrow in the inset of Fig.3(c)).  Figures 3(a1) and 3(a2) plot the dispersion maps for MI36K symmetrized about $E_{\rm F}$ and the 2nd derivative of those \cite{SM}.  Notably the spectra above and at $T_{\rm MI} \sim$36K are virtually identical, showing that there is no significant precursor of the MI transition.  At temperature is dropped below $T_{\rm MI}$, a gap opens at the Fermi node. This variation is also seen in the ARPES mapping at $E_{\rm F}$
 along a $k_x-k_y$ sheet (red plane in the inset of Fig.3(c));
the strong intensity at $\Gamma$ coming from the Fermi node (Fig. 3(d1), $T$=50K) vanishes below  $T_{\rm MI}$ (Fig. 3(d2), $T$=11K).  The band dispersion, determined from the peak/shoulder of the EDC, shown in Fig.3(c), also reflects the continuous opening of a gap below $T_{\rm MI}$. These observations are consistent with a mean-field quasiparticle dispersion, in which the gap is directly controlled by the antiferromagnetic order parameter.

However, the EDCs themselves reflect strong correlations. In Figs. 3(b1) and 3(b2), the spectra for $T=47$K and 1K corresponding to the images in Fig.3(a1) are plotted. The electronic structure in the metallic phase (Fig. 3(b1)) consists of well-defined quasiparticle peaks (red bars).  In contrast, the insulating phase (Fig. 3(b2)) shows a non-dispersive flat band, and only the broad spectra lacking long-lived elections are detected, pointing to correlation-induced Mott localization.  This circumstance contrasts to the insulating phase of Sr$_2$IrO$_4$ which shows relatively sharp, clearly dispersing spectra \cite{Kim:2008gi}.  

We investigate this further through the  detailed variation of spectral-shape at  $\Gamma$.  Figure 4(a1-c1) show the symmetrized EDCs from above to below $T_{\rm MI}$ for the three samples (MI19K, MI25K, and MI36K).
In the symmetrized EDCs the gap is reflected in two split peaks (black arrows) below $T_{\rm MI}$. 
Please note that the tracing of peak positions slightly underestimate the ``real" onset temperature of gap opening, 
especially in 3D materials with broadened spectra 
due to the imperfect sample surface and $k_z$ broadening of ARPES.
Nevertheless, the persistance of quasiparticle peaks below but near $T_{\rm MI}$ and their shift with temperature is in accord with a Slater picture \cite{Slater}. 
This is fully consistent with the recent discovery of an insulator to metal transition driven by external 
magnetic field in Nd$_2$Ir$_2$O$_7$ \cite{Tian_NP,MagTrans_Ueda}, revealing that the destruction of AIAO magnetic order restores the metallic transport.  

However, the data shows that the quasiparticle peak is significantly suppressed as temperature is further decreased, and it totally disappears at the lowest temperature, leaving only a broad spectrum.  The abnormal variation of the quasiparticle peak is also visible in the raw EDCs (Fig. 4(d1)).  While a tiny peak survives in MI19K (see Fig. 4(a1)) even at $T=$1K, it is attributable to small carrier doping in the insulating ground state due to the off-stoichiometry in the crystal \cite{Okada_NP}.  The peak suppression is examined in Fig. 4(a2-d2) in more detail, where the spectra of Fig. 4(a1-d1) normalized to the intensities around -0.3eV are overlapped with each other. The spectral weight at $E_{\rm F}$ is gradually depleted on cooling down to the lowest temperature. This feature is more clearly demonstrated in Fig. 4(e) by plotting a spectral loss near $E_{\rm F}$ ($W_{\rm loss}$) associated with the gap formation; we subtract the spectral intensities at the highest temperature from those at lower temperatures, and estimate a negative area in the difference spectra for each temperature (see the inset of Fig. 4(e)).   The pseudogap-like spectral loss quantifies the crossover from the Slater-like mean field behavior near $T_{\rm MI}$ to the Mott regime at the lowest temperatures.   The fact that the gap, extracted in Fig. 4(f), reaches $\sim$30-40 meV at low temperature (comparable to the optical gap \cite{Ueda:2012ee}), is as large as the measured bandwidth, indicates the strong coupling limit, and may be responsible for this crossover.  % The temperature dependence of the gap is distinctly different from that in the planar iridate Nd$_2$IrO$_3$, \cite{Na2IrO3_Damascelli} which remains unchanged across the magnetic transition temperature.  
% Rather the  Nd$_2$Ir$_2$O$_7$ seems to stay in the intermediate regime between Slater-type and Mott-type insulators \cite{Arita_slater,Hsieh:2012ch,Na2IrO3_slater,Haskel_PRL}.

Theory predicts that the Weyl points may migrate from the $\Gamma$ point to the zone boundary and annihilate when the order parameter becomes too large \cite{WitczakKrempa:2012cd,WitczakKrempa:2014hz,Moon:2013ha}, which may explain their absence in low temperature Nd$_2$Ir$_2$O$_7$.   One might therefore contemplate their re-appearance at intermediate temperatures just below $T_{\rm MI}$, where the gap is smaller and quasiparticles are still well-defined.    However, no indication of Weyl points at intermediate temperatures was found in the present ARPES measurements.  Apart from the difficulty of locating incommensurate temperature-dependent features in ARPES, the progressive destruction of quasiparticles we observed may be another reason for this.  % We can only hope to observe well-defined Weyl fermions in a narrow region of temperature in which the quasiparticles survive. 
We leave a dedicated search for Weyl points just below $T_{\rm MI}$, perhaps using spin-resolved ARPES, for future work.  

In conclusion, we carried out the first ARPES investigation of the MI transition of a three dimensional iridate.  We observe a quadratic Fermi node in the metallic state of Nd$_2$Ir$_2$O$_7$ very similar to that of Pr$_2$Ir$_2$O$_7$.    Upon lowering temperature below $T_{\rm MI}\sim$30K, we found a drastic variation in the spectral shape, with a gradual opening of a gap and accompanying suppression of the quasiparticle peak. At the lowest achievable temperature of 1K,  quasiparticles are completely suppressed and a dispersionless spectral edge is observed.   The results indicate a crossover from a Slater-like mean-field effective band insulator just below $T_{\rm MI}$ to a Mott-like insulator with localized electrons at the lowest temperature.

 We thank H. Shinaoka and E.-G. Moon for fruitful discussions and constructive comments, and D. Evtushinsky and 
E. Rienks for technical supports at 1$^3$ beamline in BESSY-II. 
 This work was supported by JSPS KAKENHI (Nos. 24740218, 25220707, and 25707030), by the Photon and Quantum Basic Research
Coordinated Development Program from MEXT,  by CREST, Japan Science and Technology Agency,  Grants-in-Aid for Scientific Research (No. 25707030), by Grants-in-Aids
for Scientific Research on Innovative Areas (15H05882, 15H05883), and Program for Advancing Strategic International Networks to Accelerate the Circulation of Talented Researchers (No. R2604) from the Japanese Society for the Promotion of Science.
L.B. is  supported by DOE grant DE-FG02-08ER46524. 

M. N. and T. K. contributed equally to this work.

%\bibliography{All_papers}

\end{document}

\title{Supplemental material for Slater to Mott crossover in the metal to insulator transition of Nd$_2$Ir$_2$O$_7$}

\author{M.~Nakayama}
\affiliation{ISSP, University of Tokyo, Kashiwa, Chiba 277-8581, Japan}

\author{Takeshi Kondo}
\email{kondo1215@issp.u-tokyo.ac.jp}
\affiliation{ISSP, University of Tokyo, Kashiwa, Chiba 277-8581, Japan}

\author{Z.~Tian}
\affiliation{ISSP, University of Tokyo, Kashiwa, Chiba 277-8581, Japan}

\author{J.J.~Ishikawa}
\affiliation{ISSP, University of Tokyo, Kashiwa, Chiba 277-8581, Japan}

\author{M.~Halim}
\affiliation{ISSP, University of Tokyo, Kashiwa, Chiba 277-8581, Japan}

\author{C.~Bareille}
\affiliation{ISSP, University of Tokyo, Kashiwa, Chiba 277-8581, Japan}

\author{W.~Malaeb}
\affiliation{ISSP, University of Tokyo, Kashiwa, Chiba 277-8581, Japan}
\affiliation{Physics Department, Faculty of Science, Beirut Arab University, Beirut, Lebanon}

\author{K.~Kuroda}
\affiliation{ISSP, University of Tokyo, Kashiwa, Chiba 277-8581, Japan}

\author{T.~Tomita}
\affiliation{ISSP, University of Tokyo, Kashiwa, Chiba 277-8581, Japan}

\author{S.~Ideta}
\affiliation{UVSOR Facility, Institute for Molecular Science, Okazaki 444-8585, Japan}

\author{K.~Tanaka}
\affiliation{UVSOR Facility, Institute for Molecular Science, Okazaki 444-8585, Japan}

\author{M.~Matsunami}
\affiliation{Toyota Technological Institute, Nagoya 468-8511, Japan}

\author{S.~Kimura}
\affiliation{Graduate School of Frontier Biosciences, Osaka University, Suita, Osaka, 565-0871, Japan }

\author{N.~Inami}
\affiliation{Institute of Materials Structure Science, High Energy Accelerator Research Organization (KEK),  Tsukuba, Ibaraki 305-0801, Japan}

\author{K.~Ono}
\affiliation{Institute of Materials Structure Science, High Energy Accelerator Research Organization (KEK),  Tsukuba, Ibaraki 305-0801, Japan}

\author{H.~Kumigashira}
\affiliation{Institute of Materials Structure Science, High Energy Accelerator Research Organization (KEK),  Tsukuba, Ibaraki 305-0801, Japan}

\author{L.~Balents}
\affiliation{Kavli Institute for Theoretical Physics, Santa Barbara, California 93106, USA}

\author{S.~Nakatsuji}
\affiliation{ISSP, University of Tokyo, Kashiwa, Chiba 277-8581, Japan}
\affiliation{CREST, Japan Science and Technology Agency (JST), 4-1-8 Honcho Kawaguchi, Saitama 332-0012, Japan}

\author{S.~Shin}
\affiliation{ISSP, University of Tokyo, Kashiwa, Chiba 277-8581, Japan}

\date{\today}

\maketitle

%\setstretch{1.1}
\section{Energy distribution  curves (EDCs) used for  the symmetrization  analysis}

\begin{figure*}
\includegraphics[width=6.5in]{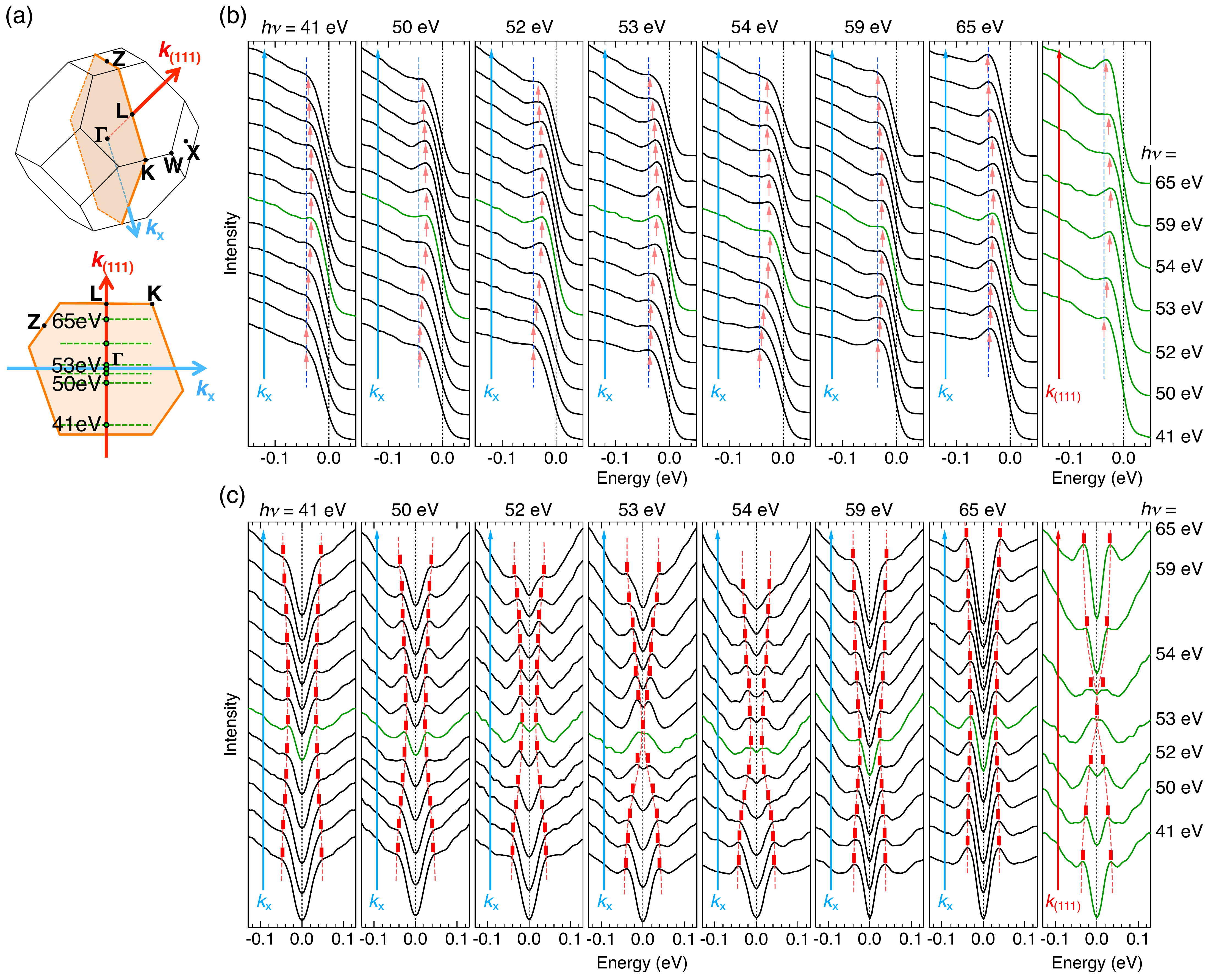}
\caption{(a) Brillouin zone of Nd$_2$Ir$_2$O$_7$ with measured momentum cuts at different photon energies (dashed lines).  (b) EDCs measured along the $k_x$ direction defined in (a). The used photon energies are described on top of each panel, and the corresponding momentum cuts are indicated with dashed lines in (a). On the right panel, the EDCs at  $k_x$=0 (or along $\Gamma$-L; green circles in (a)) are extracted. 
The energy positions of spectral peaks are marked by arrows. The dashed blue lines are a guide to the eyes to confirm the energy dispersion in the data. (c) The symmetrized EDCs corresponding to the data in (b). The energy dispersions are indicated by red bars and dashed curves.}
\label{FigS1}
\end{figure*}

In the main paper, we show the energy distribution curves (EDCs) symmetrized about $E_{\rm F}$ to remove the cut-off effect on the spectral shape by  Fermi function. Here we present the row EDCs used for the  analysis (Fig.~\ref{FigS1}).

 Figure~\ref{FigS1} shows the EDCs in the metallic phase ($T$=50K) of Nd$_2$Ir$_2$O$_7$, 
measured at various photon energies from $h\nu$ = 41 eV to 65 eV. The measured momentum cuts (dashed lines in  Fig.~\ref{FigS1}(a)) are located in the $k_x-k_{(111)}$ plane crossing $\Gamma$, Z, L, and K.  As marked by magenta arrows in Fig.~\ref{FigS1}(a), the quasiparticle peaks clearly disperse along $k_x$,  and that at $k_x=0$ (green curve in each panel) most approaches $E_{\rm F}$ at $h\nu$ = 53 eV as examined in the right panel of Fig.~\ref{FigS1}(b). The symmetrized EDCs of  Fig.~\ref{FigS1}(b) are plotted in Fig.~\ref{FigS1}(c). It is clearly seen that the quadratic dispersion approaches $E_{\rm F}$ with increasing the photon energy, touches $E_{\rm F}$ at $h\nu$ = 53 eV, and  move away from $E_{\rm F}$ again with a further increase of $h\nu$. 
This behavior is more clearly demonstrated in the right panel of  Fig.~\ref{FigS1}(c) [same as Fig. 2(d2) in the main paper], where the curves at $k_x=0$ are extracted. 
%\\\\

\section{Agreement between the results obtained with symmetrized spectra and FD-divided spectra}

In the main paper,  we mainly use the symmetrization method to remove the Fermi cut-off effect from the ARPES spectra (EDCs);  the EDCs are flipped about $E_F$ and added to the original curves. 
This technique is widely used to visualize the opening of a gap, and also to determine the Fermi crossing point ($k_F$ point). 
The method is easy to use and applicable even for the low temperature spectra with negligible intensity above $E_F$, thus 
it has been commonly used for a gap estimation in the superconducting and density-wave materials. 
A weakness of this method, however, is that it requires the particle-hole symmetry, and its validity is not for sure in unknown materials like Nd$_2$Ir$_2$O$_7$.   

The more straightforward way of eliminating the Fermi cut-off effect would be to divide the EDCs by the Fermi function (FD) at measured temperature. 
For a realistic analysis, the Fermi function actually used is convoluted with the experimental energy resolution. 
The FD-division method, though conceptually straightforward, 
also has a difficulty to be applied for the data. First of all, the high temperature data are required, since it utilizes
 thermally populated intensities above $E_F$. Moreover, 
  high quality data with a high signal/noise ratio are essential in order to reliably reproduce 
 the spectral shape close to $E_F$. 
Therefore, the best way to extract the band dispersion close to $E_F$ is to 
 apply both analyses of the symmetrization and FD-division methods to the same data, and confirm a consistency between the two results. 
 
\begin{figure*}
\includegraphics[width=2.8in]{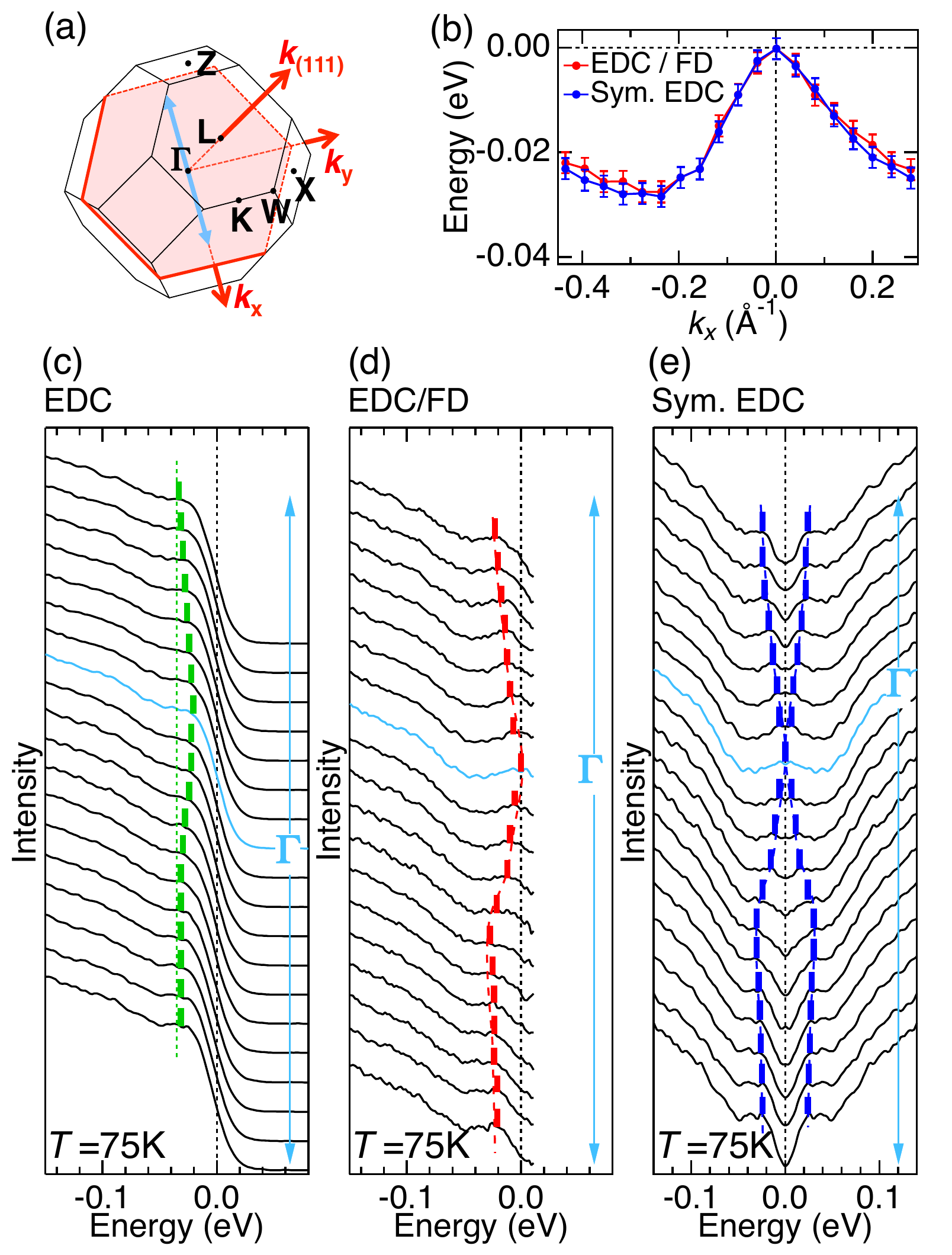}
\caption{(a) Brillouin zone of Nd$_2$Ir$_2$O$_7$ with the measured momentum cut crossing $\Gamma$ (a light blue arrow). (b) The Fermi node state consistently obtained with two different methods: the symmetrization and the FD-division. (c) The EDCs along a momentum cut across $\Gamma$ (a light blue arrow) measured at $T=75$K.  The dashed blue line is a guide to the eyes to confirm the energy dispersion in the data. 
(d) EDCs in (c) divided by the Fermi function at the measured temperature ($T=75$K). (e) EDCs in (c) symmetrized about $E_F$. The peak energies of spectra are marked by colored bards and dashed curves. 
}
\label{FigS2}
\end{figure*}

 In Fig.~\ref{FigS2}, we demonstrate that the results obtained with these two methods agree with each other,
 pointing to the realization of the quadratic Fermi node state in the metallic phase of Nd$_2$Ir$_2$O$_7$. 
Figure~\ref{FigS2}(c) plot the EDCs along a momentum cut across $\Gamma$ (a light blue arrow in  Fig.~\ref{FigS2}(a)).
Here we use the data at $T=75$K, which fulfill the following two conditions: the surviving of quasiparticles  
and the sufficient intensities in the thermally populated spectral weight close to $E_F$. 
The EDCs in Fig.~\ref{FigS2}(c) divided by the Fermi function and those symmetrized about $E_F$ are 
exhibited in  Figs.~\ref{FigS2}(d) and \ref{FigS2}(e), respectively. 
The effect of Fermi cut-off near $E_F$
is removed by these treatments, and importantly, the dispersion of  spectral peaks (color bars and dashed curves) 
touches $E_F$ at the $\Gamma$ point in both the cases.  The results are more clearly demonstrated in  Fig.~\ref{FigS2}(b), where the peak positions in  Figs.~\ref{FigS2}(d) and \ref{FigS2}(e)
are extracted. We find that the plots obtained by using the symmetrization and the FD-division methods (blue and red circles, respectively) almost
perfectly match with each other, both showing the Fermi node state. The consistency between the two different analyses validates our conclusion.  
%\\\\

\section{variation of Band topology through the METAL-INSULATOR TRANSITION}

In Fig.3 of the main paper, we show the temperature evolution of band dispersion only for the sample MI36K with the highest transition temperature ($T_{\rm MI}$) of 36K among three samples we used. Here we also exhibit it for the other samples of MI19K and MI25K with the lower $T_{\rm MI}$ of 19K and 25K, respectively, and compare the results among the three samples. 

\begin{figure*}
\includegraphics[width=6in]{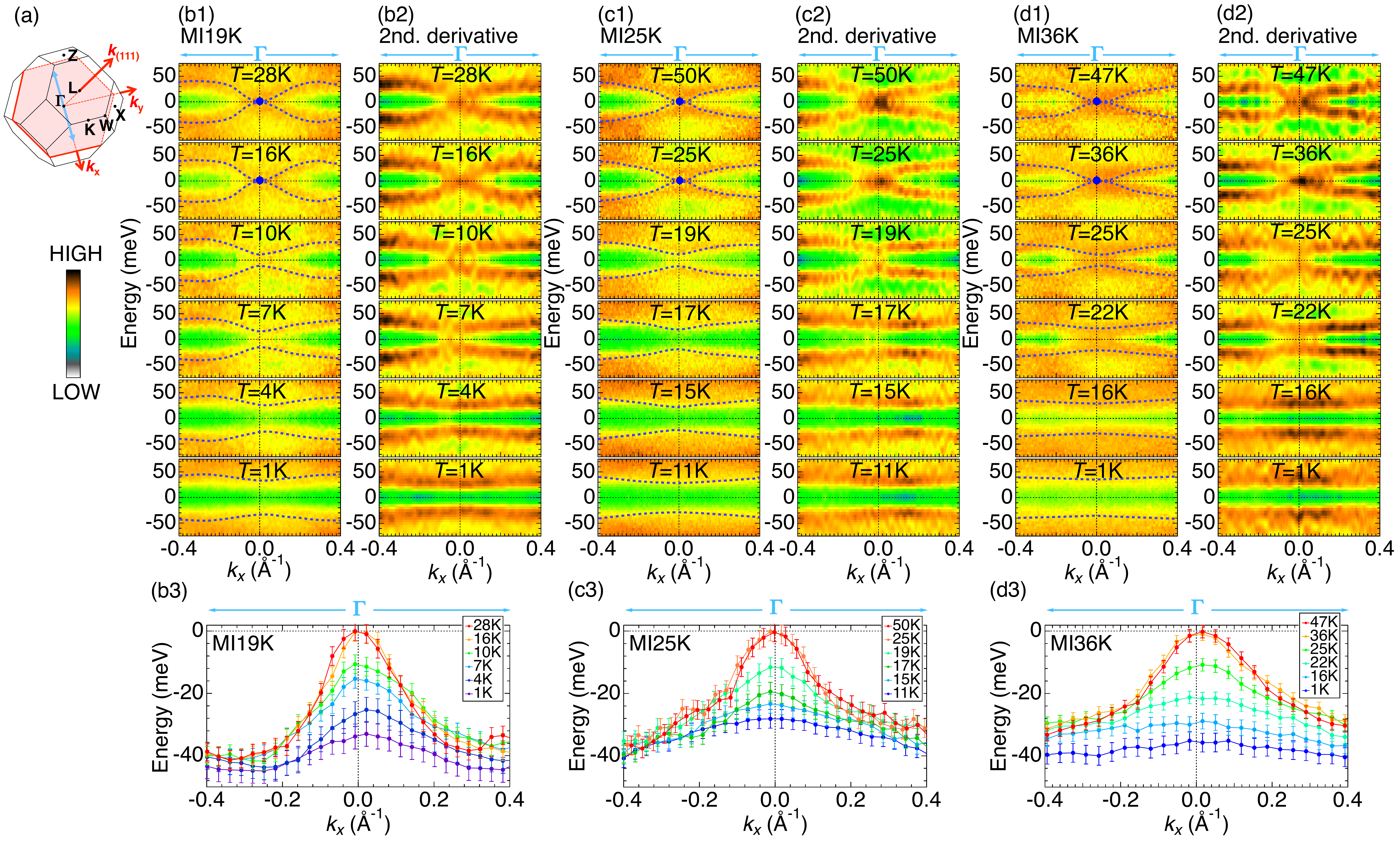}
\caption{(a) Brillouin zone of Nd$_2$Ir$_2$O$_7$ with the measured momentum cut crossing $\Gamma$ (a light blue arrow). (b-d) Temperature evolution of ARPES dispersion measured along a momentum cut across $\Gamma$ (a light blue arrow in (a)) for three samples of MI19K, MI25K, and MI36K, respectively. 
The left panels (b1-d1) show the ARPES images symmetrized about $E_F$. In the right panels (b2-d2), the 2nd derivatives of the same images as (b1-d1) are plotted. (b3-d3) The energy dispersion determined from the peak energies of spectra for MI19K, MI25K, and MI36K, respectively. 
}
\label{FigS3}
\end{figure*}

 Figure \ref{FigS3}(b1) and \ref{FigS3}(c1) show ARPES dispersion maps measured along a momentum cut across $\Gamma$ (a light blue arrow in Fig. \ref{FigS3}(a)) for MI19K and MI25K, respectively. Each image is symmetrized about $E_F$ to visualize whether or not a gap is open. This is more clearly exhibited in  Fig. \ref{FigS3}(b2) and \ref{FigS3}(c2), where we plot the 2nd derivative of each left image in the Figs. \ref{FigS3}(b1) and  \ref{FigS3}(c1), respectively. The Fermi node sate with a band touching at $E_F$ is observed in the metallic phase above $T_{\rm MI}$.  Upon cooling below the $T_{\rm MI}$, the band structure becomes gapped, and the magnitude of the energy gap increases with a decrease of temperature. The band topology is eventually turned to be almost flat at the lowest temperature, meaning that the electrons are well localized with a negligible hopping. 
 We summarize the anomalous behaviors in Figs.~\ref{FigS3}(b3) and \ref{FigS3}(c3) by plotting the energy dispersions determined by the peak positions of spectra for MI19K and MI25K, respectively. For a comparison, we also exhibit the results of MI36K in Fig. \ref{FigS3}(d) (same data as in Fig. 3 of the main paper). While the band topology of MI36K becomes completely flat at the lowest temperature ($T=1$K), the weak dispersion seems to remain in MI19K and MI25K even at the lowest measured temperature. The lowest achievable temperature was only $\sim10$K 
 during the experiment for MI25K conducted in BL28A of the KEK facility, hence the complete localization of electrons (or flat band) is possibly realized in the ground state of MI25K as that of MI36K. 
 In contrast, the data for MI19K show a clear dispersion even at $T=1$K, which is available in the 1$^3$ beamline of BESSY-II. This indicates that the itinerancy of conduction electrons remains in the ground state of MI19K. The relevant signature is a tiny peak surviving even at at $T=1$K (see Fig. 4(a1)), that is the lowest achievable temperature in ARPES. 
  
The  variation in the spectral feature among crystal pieces should come from a different degree of the off-stoichiometry (Nd$_{2-x}$Ir$_{2+x}$O$_7$ ), which dopes a slight carrier to the crystals. In fact, the electron-probe microanalysis (EPMA) estimates the Ir/Nd ratio of approximately 1$\%$ and 2$\%$ in our single crystals with zero-field $T_{\rm MI}$ of $\sim$25K and $\sim$20K, respectively. 
A tiny peak observed in MI19K  at  $T=$1K (see Fig. 4(a1)) is therefore attributed to a small carrier doping to the Mott insulating ground state, caused by the off-stoichiometry \cite{Okada_NP}.
%\\

%\bibliography{All_papers}

%%%%%%%%%%%%%%%%%%%%%%
%%%%%%%%%%%%%%%%%%%%%%
%\clearpage
%\newpage 

%%%%%%%%%%%%%%%%%%%%%%
%%%%%%%%%%%%%%%%%%%%%%
%\cellarage
%\newpage 

%%%%%%%%%%%%%%%%%%%%%%
%%%%%%%%%%%%%%%%%%%%%%
%\clearpage
%\newpage 